\documentstyle [12pt]{article}
\begin{document}
\title{ Scale Invariance, Mass and Cosmology 
\protect\\  } \author{E.I. Guendelman  
\\{\it Physics Department, Ben-Gurion University, Beer-Sheva
84105, Israel}}

\maketitle
\bigskip

\begin{abstract}
The possibility of mass in the context of scale-invariant, generally
covariant theories, is discussed. The realizations of scale invariance
which are considered, are in the context of a gravitational theory where
the action, in the first order formalism, is of the form $S =
\int L_{1} \Phi d^4x$ + $\int L_{2}\sqrt{-g}d^4x$ where $\Phi$ is a
density built out of degrees of freedom independent of gravity, which we
call the "measure fields". For global scale invariance, a "dilaton"
$\phi$ has to be introduced, with non-trivial potentials $V(\phi)$ =
$f_{1}e^{\alpha\phi}$ in $L_1$ and $U(\phi)$ = $f_{2}e^{2\alpha\phi}$ in
$L_2$. This leads to non-trivial mass generation and potential for
$\phi$. Mass terms for an arbitrary matter field can appear in a scale
invariant form both in $L_1$ and in $L_2$ where they are coupled to
different exponentials of the field $\phi$. Implications of these results 
for cosmology having in mind in particular inflationary scenarios, models 
of the late universe and modified gravitational theories are discussed.
\end{abstract}

\section{Introduction}

The concept of scale invariance appears as an attractive possibility for a
fundamental symmetry of nature. In its most naive realizations, such a
symmetry is not a viable symmetry, however, since nature seems to have
chosen some typical scales.

Here we will find that scale invariance can nevertheless be incorporated
into realistic, generally covariant field theories. However, scale
invariance has to be discussed in a more general framework than that of
standard generally relativistic theories, where we must allow in the
action, in addition to the
ordinary measure of integration $\sqrt{-g}d^{4}x$, another one$^1$,
$\Phi d^{4}x$, where $\Phi$ is a density built out of degrees of freedom
independent of that of $g_{\mu\nu}$. To achieve global scale invariance,
also a "dilaton" $\phi$ has to be introduced$^2$.

As will be discussed, a potential consistent with scale invariance can
appear for the $\phi$ field. Such a potential has a shape which makes it
suitable for the satisfactory realization of an inflationary
scenario$^3$ of the improved type$^4$. Alternatively, it can be of use
in a slowly rolling $\Lambda-$ scenario for the late universe$^5$.

Finally, we also discuss how scale invariant mass terms, which lead to 
phenomenologically acceptable dynamics, can be introduced into the theory.
We discuss some properties of such types of mass terms and
their implication for the early universe inflationary cosmology,
for the cosmology of a late universe filled with matter and for the 
possibility of obtaining modified gravitational dynamics.

\section{The Non Gravitating Vacuum Energy (NGVE) Theory. Strong and Weak Formulations.}

	When formulating generally covariant Lagrangian formulations of
gravitational theories, we usually consider the form
\begin{equation}
S_{1} = \int{L}\sqrt{-g} d^{4}x, g =  det g_{\mu\nu}
\end{equation}

	As it is well known, $d^{4}x$ is not a scalar but the combination
$\sqrt{-g} d^{4} x$ is a scalar. Inserting $\sqrt{-g}$,
which has the transformation properties of a
density, produces a scalar action (1), provided $L$ is a scalar.

	One could use nevertheless other objects instead of
$\sqrt{-g}$, provided
they have the same transformation properties and achieve in this way a
different generally covariant formulation.

	For example, given 4-scalars $\varphi_{a}$ (a = 
1,2,3,4), one can construct the density
\begin{equation} 
\Phi =  \varepsilon^{\mu\nu\alpha\beta}  \varepsilon_{abcd}
\partial_{\mu} \varphi_{a} \partial_{\nu} \varphi_{b} \partial_{\alpha}
\varphi_{c} \partial_{\beta} \varphi_{d}  
\end{equation}
and consider instead of (1) the action$^1$
\begin{equation}
S_{2} =  \int L \Phi d^{4} x.	
\end{equation}
L is again some scalar, which contains the curvature (i.e. the
gravitational contribution) and a matter contribution, as it is standard also
in (1).

	In the action (3) the measure carries degrees of freedom
independent of that of the metric and that of the matter fields. The most
natural and successful formulation of the theory is achieved when the
connection coefficients are also treated as an independent degrees  of freedom.
 This is what is usually referred to as the first order formalism.

	One can notice that $\Phi$ is the total derivative of something,
for
example, one can write
\begin{equation}
\Phi = \partial_{\mu} ( \varepsilon^{\mu\nu\alpha\beta}
\varepsilon_{abcd} \varphi_{a}
  \partial_{\nu} \varphi_{b}
\partial_{\alpha}
\varphi_{c} \partial_{\beta} \varphi_{d}). 
\end{equation}

	This means that a shift of the form
\begin{equation}	
		L \rightarrow  L  +  constant	
\end{equation}
just adds the integral of a total divergence to the action (3) and it does
not affect therefore the equations of motion of the theory. The same
shift, acting on (1) produces an additional term which gives rise to a
cosmological constant.  
Since the constant part of L does not affect the equations of motion resulting
 from the action (3), this
theory is called the Non Gravitating Vacuum Energy (NGVE) Theory$^1$.

	One can generalize this structure and allow both geometrical
objects to enter the theory and consider
\begin{equation}
S_{3} = \int L_{1} \Phi  d^{4} x  +  \int L_{2} \sqrt{-g}d^{4}x		
\end{equation}

	Now instead of  (5),  the shift symmetry can be applied only on
$L_{1}$ 
($L_{1} \rightarrow L_{1}$ + constant). Since the structure has been
generalized, we call
this formulation the weak version of the NGVE - theory. Here $L_{1}$ and
$L_{2}$ are
$\varphi_{a}$  independent.

	There is a good reason not to consider mixing of  $\Phi$ and
$\sqrt{-g}$ , like
for example using
\begin{equation}
\frac{\Phi^{2}}{\sqrt{-g}} 
\end{equation}		 	

this is because (6) is invariant (up to the integral of a total
divergence) under the infinite dimensional symmetry
\begin{equation}
\varphi_{a} \rightarrow \varphi_{a}  +  f_{a} (L_{1})	
\end{equation}
where $f_{a} (L_{1})$ is an arbitrary function of $L_{1}$ if $L_{1}$ and
$L_{2}$ are $\varphi_{a}$
independent. Such symmetry (up to the integral of a total divergence) is
absent if mixed terms (like (7)) are present.  Therefore (6) is considered
for the case when no dependence on the measure fields (MF) appears in
$L_{1}$ or $L_{2}$.

	In this paper  we will see that the existence of two independent
measures of integrations as in (6) allows new realizations of global scale
invariance with most interesting consequences when the results are viewed
from the point of view of cosmology.

\section{The Action Principle for a Scalar Field in the Weak NGVE - Theory}

	We will study now the dynamics of a scalar field $\phi$ interacting
with gravity as given by the following action
\begin{equation}
S_{\phi} =  \int L_{1} \Phi d^{4} x  +  \int L_{2} \sqrt{-g}   d^{4} x
\end{equation}
\begin{equation}
L_{1} = \frac{-1}{\kappa} R(\Gamma, g) + \frac{1}{2} g^{\mu\nu}
\partial_{\mu} \phi \partial_{\nu} \phi - V(\phi) 
\end{equation}	
\begin{equation}
L_{2} = U(\phi)
\end{equation}
\begin{equation}	
R(\Gamma,g) =  g^{\mu\nu}  R_{\mu\nu} (\Gamma) , R_{\mu\nu}
(\Gamma) = R^{\lambda}_{\mu\nu\lambda}
\end{equation}
\begin{equation}
R^{\lambda}_{\mu\nu\sigma} (\Gamma) = \Gamma^{\lambda}_
{\mu\nu,\sigma} - \Gamma^{\lambda}_{\mu\sigma,\nu} +
\Gamma^{\lambda}_{\alpha\sigma}  \Gamma^{\alpha}_{\mu\nu} -
\Gamma^{\lambda}_{\alpha\nu} \Gamma^{\alpha}_{\mu\sigma}.	 
\end{equation}

	In the variational principle $\Gamma^{\lambda}_{\mu\nu},
g_{\mu\nu}$, the measure fields scalars
$\varphi_{a}$ and the  scalar field $\phi$ are all to be treated
as independent
variables although the variational principle may result in equations that
allow us to solve some of these variables in terms of others.

\section{Global Scale Invariance}

	If we perform the global scale transformation ($\theta$ =
constant) 
\begin{equation}
g_{\mu\nu}  \rightarrow   e^{\theta}  g_{\mu\nu}	
\end{equation}
then (9) is invariant provided  $V(\phi)$ and $U(\phi)$ are of the
form  
\begin{equation}
V(\phi) = f_{1}  e^{\alpha\phi},  U(\phi) =  f_{2}
e^{2\alpha\phi}
\end{equation}
and $\varphi_{a}$ is transformed according to
\begin{equation}
\varphi_{a}   \rightarrow   \lambda_{a} \varphi_{a}  
\end{equation}
(no sum on a) which means
\begin{equation}
\Phi \rightarrow \biggl(\prod_{a} {\lambda}_{a}\biggr) \Phi \\ \equiv \lambda 
\Phi	 \end{equation}
such that
\begin{equation} 
\lambda = e^{\theta}
\end{equation}	
and	 
\begin{equation}
\phi \rightarrow \phi - \frac{\theta}{\alpha}.     	
\end{equation}

In this case we call the scalar field $\phi$ needed to implement scale 
invariance "dilaton".

\section{The Equations of Motion}

	We will now work out the equations of motion for arbitrary choice
of $V(\phi)$ and $U(\phi)$. We study afterwards the choice (15) which
allows us to
obtain the results for the scale invariant case and also to see what
differentiates this from the choice of arbitrary $U(\phi)$ and  $V(\phi)$ 
in a very
special way.

	Let us begin by considering the equations which are obtained from
the variation of the fields that appear in the measure, i.e. the
$\varphi_{a}$
fields. We obtain then  
\begin{equation}		
A^{\mu}_{a} \partial_{\mu} L_{1} = 0   	
\end{equation}
where  $A^{\mu}_{a} = \varepsilon^{\mu\nu\alpha\beta}
\varepsilon_{abcd} \partial_{\nu} \varphi_{b} \partial_{\alpha}
\varphi_{c} \partial_{\beta} \varphi_{d}$. Since it is easy to
check that  $A^{\mu}_{a} \partial_{\mu} \varphi_{a^{\prime}} =
\frac{\delta aa^{\prime}}{4} \Phi$, it follows that 
det $(A^{\mu}_{a}) =\frac{4^{-4}}{4!} \Phi^{3} \neq 0$ if $\Phi\neq 0$.
Therefore if $\Phi\neq 0$ we obtain that $\partial_{\mu} L_{1} = 0$,
 or that
\begin{equation}
L_{1} = \frac{-1}{\kappa} R(\Gamma,g) + \frac{1}{2} g^{\mu\nu}
\partial_{\mu} \phi \partial_{\nu} \phi - V = M	 
\end{equation}
where M is constant.

	Let us study now the equations obtained from the variation of the
connections $\Gamma^{\lambda}_{\mu\nu}$.  We obtain then
\begin{equation}
-\Gamma^{\lambda}_{\mu\nu} -\Gamma^{\alpha}_{\beta\mu}
g^{\beta\lambda} g_{\alpha\nu}  + \delta^{\lambda}_{\nu}
\Gamma^{\alpha}_{\mu\alpha} + \delta^{\lambda}_{\mu}
g^{\alpha\beta} \Gamma^{\gamma}_{\alpha\beta}
g_{\gamma\nu}\\ - g_{\alpha\nu} \partial_{\mu} g^{\alpha\lambda}
+ \delta^{\lambda}_{\mu} g_{\alpha\nu} \partial_{\beta}
g^{\alpha\beta}         \\
 - \delta^{\lambda}_{\nu} \frac{\Phi,_\mu}{\Phi}
+ \delta^{\lambda}_{\mu} \frac{\Phi,_\nu}{\Phi} =  0	
\end{equation}
If we define $\Sigma^{\lambda}_{\mu\nu}$    as
$\Sigma^{\lambda}_{\mu\nu} =
\Gamma^{\lambda}_{\mu\nu} -\{^{\lambda}_{\mu\nu}\}$
where $\{^{\lambda}_{\mu\nu}\}$   is the Christoffel symbol, we
obtain for $\Sigma^{\lambda}_{\mu\nu}$ the equation 
\begin{equation}
	-  \sigma, _{\lambda} g_{\mu\nu} + \sigma, _{\mu}
g_{\nu\lambda} - g_{\nu\alpha} \Sigma^{\alpha}_{\lambda\mu}
-g_{\mu\alpha} \Sigma^{\alpha}_{\nu \lambda}
+ g_{\mu\nu} \Sigma^{\alpha}_{\lambda\alpha} +
g_{\nu\lambda} g_{\alpha\mu} g^{\beta\gamma} \Sigma^{\alpha}_{\beta\gamma}
= 0 
\end{equation}		 
where  $\sigma = ln \chi, \chi \equiv \frac{\Phi}{\sqrt{-g}}$.
      	
	The general solution of (23) is 
\begin{equation}
\Sigma^{\alpha}_{\mu\nu} = \delta^{\alpha}_{\mu}
\lambda,_{\nu} + \frac{1}{2} (\sigma,_{\mu} \delta^{\alpha}_{\nu} -
\sigma,_{\beta} g_{\mu\nu} g^{\alpha\beta})
\end{equation}
where $\lambda$ is an arbitrary function due to the $\lambda$ - symmetry
of the
curvature$^6$  $R^{\lambda}_{\mu\nu\alpha} (\Gamma)$,
\begin{equation}
\Gamma^{\alpha}_{\mu\nu} \rightarrow \Gamma^{\prime \alpha}_{\mu\nu}
 = \Gamma^{\alpha}_{\mu\nu} + \delta^{\alpha}_{\mu}
Z,_{\nu}
\end{equation} 
Z  being any scalar (which means $\lambda \rightarrow \lambda + Z$).
  
	If we choose the gauge $\lambda = \frac{\sigma}{2}$, we obtain
\begin{equation}
\Sigma^{\alpha}_{\mu\nu} (\sigma) = \frac{1}{2} (\delta^{\alpha}_{\mu}
\sigma,_{\nu} +
 \delta^{\alpha}_{\nu} \sigma,_{\mu} - \sigma,_{\beta}
g_{\mu\nu} g^{\alpha\beta}).
\end{equation}

	Considering now the variation with respect to $g^{\mu\nu}$, we
obtain
\begin{equation}	 	
\Phi (\frac{-1}{\kappa} R_{\mu\nu} (\Gamma) + \frac{1}{2} \phi,_{\mu}
\phi,_{\nu}) - \frac{1}{2} \sqrt{-g} U(\phi) g_{\mu\nu} = 0
\end{equation}
Solving for $R = g^{\mu\nu} R_{\mu\nu} (\Gamma)$  and introducing in
(21), we obtain a contraint,
\begin{equation}
M + V(\phi) - \frac{2U(\varphi)}{\chi} = 0
\end{equation}
that allows us to solve for $\chi$,
\begin{equation}
\chi = \frac{2U(\phi)}{M+V(\phi)}.
\end{equation}

	To get the physical content of the theory, it is convenient to go
to the Einstein conformal frame where 
\begin{equation}
\overline{g}_{\mu\nu} = \chi g_{\mu\nu}		    
\end{equation}
and $\chi$  given by (29). In terms of $\overline{g}_{\mu\nu}$   the non
Riemannian contribution $\Sigma^{\alpha}_{\mu\nu}$
disappears from the equations, which can be written then in the Einstein
form ($R_{\mu\nu} (\overline{g}_{\alpha\beta})$ =  usual Ricci tensor)
\begin{equation}
R_{\mu\nu} (\overline{g}_{\alpha\beta}) - \frac{1}{2} 
\overline{g}_{\mu\nu}
R(\overline{g}_{\alpha\beta}) = \frac{\kappa}{2} T^{eff}_{\mu\nu}
(\phi)	 	
\end{equation}
where
\begin{equation}	 
T^{eff}_{\mu\nu} (\phi) = \phi_{,\mu} \phi_{,\nu} - \frac{1}{2} \overline
{g}_{\mu\nu} \phi_{,\alpha} \phi_{,\beta} \overline{g}^{\alpha\beta}
+ \overline{g}_{\mu\nu} V_{eff} (\phi)
\end{equation}
and 	
\begin{equation}
V_{eff} (\phi) = \frac{1}{4U(\phi)}  (V+M)^{2}.
\end{equation}
	
	In terms of the metric $\overline{g}^{\alpha\beta}$ , the equation
of motion of the Scalar
field $\phi$ takes the standard General - Relativity form
\begin{equation}
\frac{1}{\sqrt{-\overline{g}}} \partial_{\mu} (\overline{g}^{\mu\nu} 
\sqrt{-\overline{g}} \partial_{\nu}
\phi) + V^{\prime}_{eff} (\phi) = 0.
\end{equation} 

	Notice that if  $V + M = 0,  V_{eff}  =  0$ and $V^{\prime}_{eff} 
= 0$ also, provided $V^{\prime}$
is finite and $U \neq 0$ and regular there. This means the zero cosmological 
constant state is achieved without any sort of fine tuning. This is the basic
feature that characterizes the NGVE - theory and allows it to solve the
cosmological constant problem$^{1}$. It should be noticed that the equations
of motion in terms of  $\overline{g}_{\mu\nu}$ are perfectly regular at 
$V + M = 0$ although the transformation (30) is singular at this point. In terms
of the original metric $g_{\mu\nu}$ the equations do have a singularity at 
$V + M = 0$. The existence of the singular behavior in the original frame implies
the vanishing of the vacuum energy for the true vacuum state in the bar frame,
but without any singularities there.

	In what follows we will study (33) for the special case of global
scale invariance, which as we will see displays additional very special
features which makes it attractive in the context of cosmology.

	Notice that in terms of the variables $\phi$,
$\overline{g}_{\mu\nu}$, the "scale"
transformation becomes only a shift in the scalar field $\phi$, since
$\overline{g}_{\mu\nu}$ is
invariant (since $\chi \rightarrow \lambda^{-1} \chi$  and $g_{\mu\nu}
\rightarrow \lambda g_{\mu\nu}$)
\begin{equation}
\overline{g}_{\mu\nu} \rightarrow \overline{g}_{\mu\nu}, \phi \rightarrow
\phi - \frac{\theta}{\alpha}.
\end{equation}

\section{Analysis of the Scale - Invariant Dynamics}

	If $V(\phi) = f_{1} e^{\alpha\phi}$  and  $U(\phi) = f_{2}
e^{2\alpha\phi}$ as
required by scale
invariance (14), (16), (17), (18), (19), we obtain from (33)
\begin{equation}
	V_{eff}  = \frac{1}{4f_{2}}  (f_{1}  +  M e^{-\alpha\phi})^{2}	
\end{equation}

	Since we can always perform the transformation $\phi \rightarrow
- \phi$ we can
choose by convention $\alpha > O$. We then see that as $\phi \rightarrow
\infty, V_{eff} \rightarrow \frac{f_{1}^{2}}{4f_{2}} =$ const.
providing an infinite flat region. Also a minimum is achieved at zero
cosmological constant for the case $\frac{f_{1}}{M} < O$ at the point 
\begin{equation}
\phi_{min}  =  \frac{-1}{\alpha} ln \mid\frac{f_1}{M}\mid.  	
\end{equation}

	Finally, the second derivative of the potential  $V_{eff}$  at the
minimum is 
\begin{equation}
V^{\prime\prime}_{eff} = \frac{\alpha^2}{2f_2} \mid{f_1}\mid^{2} > O
\end{equation}
if
$f_{2} > O$,	 	
there are many interesting issues that one can raise here. The first one
is of course the fact that a realistic scalar field potential, with
massive excitations when considering the true vacuum state, is achieved in
a way consistent with the idea (although somewhat generalized) of scale
invariance.

	The second point to be raised is that there is an infinite region
of flat potential for $\phi \rightarrow \infty$, which makes this theory
an attractive
realization of the improved inflationary model$^{4}$.

	A peculiar feature of the potential (36), is that the integration 
constant M, provided it has the correct sign, i.e. that $f_{1}/M < 0$, 
does not affect the physics of the problem. This is because if we perform 
a shift 
\begin{equation}
\phi \rightarrow \phi + \Delta		
\end{equation}	
in the potential (36), this is equivalent to the change in the integration
constant  M
\begin{equation}
M \rightarrow M e^{-\alpha\Delta}.	
\end{equation}

	We see therefore that if we change  M in any way, without changing
the sign of M, the only effect this has is to shift the whole potential.
The physics of the potential remains unchanged, however. This is 
reminiscent of the dilatation invariance of the theory, which involves
only a shift in $\phi$  if $\overline{g}_{\mu\nu}$   is used (see eq. (35)
).

	This is very different from the situation for two generic
functions
$U(\phi)$ and 
$V(\phi)$ in (33 ). There, M appears in $V_{eff}$ as a true new parameter
that
generically changes the shape of the potential $V_{eff}$, i.e. it is
impossible
then to compensate the effect of M with just a shift. For example  M will
appear in the value of the second derivative of the potential at the
minimum, unlike what we see in eq. (38), where we see that
$V^{\prime\prime}_{eff}$ (min) is M
independent.

	In conclusion, the scale invariance of the original theory is
responsible for the non appearance (in the physics) of a certain scale,
that associated to M. However, masses do appear, since the coupling to two
different measures of $L_{1}$ and $L_{2}$ allow us to introduce two
independent
couplings  $f_{1}$ and $f_{2}$, a situation which is  unlike the
standard
formulation of globally scale invariant theories, where usually no stable
vacuum state exists.

	Notice that we have not considered all possible terms consistent
with global scale invariance. Additional terms in  $L_{2}$  of the form
$e^{\alpha\phi} R$ and \\ $e^{\alpha\phi} g^{\mu\nu} \partial_{\mu}\phi
\partial_{\nu}\phi$
 are indeed consistent with the global scale invariance
(14), (16), (17), (18), (19) but they give rise to a much more complicated
theory, which will be studied in a separate publication. There it will be
shown that for slow rolling and for $\phi \rightarrow \infty$ the
basic features of the theory
are the same as what has been studied here. Let us finish this section by
comparing the appearance of the potential $V_{eff} (\phi)$, which has
privileged
some point depending on M (for example the minimum of the potential will
have to be at some specific point), although the theory has the
"translation invariance" (35), to the physics of solitons.

	In fact, this very much resembles the appearance of solitons in a
space-translation invariant theory: The soliton solution has to be
centered at some point, which of course is not determined by the theory.
The soliton of  course breaks the space translation invariance
spontaneously, just as the existence of the non trivial potential $V_{eff}
(\phi)$
breaks here spontaneously the translations in $\phi$ space, since $V_{eff}
(\phi)$ is
not a constant. 

	Notice however, that the existence for $\phi \rightarrow \infty$,
of a flat region for
$V_{eff} (\phi)$ can be nicely described as a region where the symmetry
under
translations (35) is restored.

\section{Cosmological Applications of the Model}

	Since we have an infinite region in which $V_{eff}$ as given by
(36) is
flat $(\phi \rightarrow \infty)$, we expect a slow rolling (new
inflationary) scenario to be
viable, provided the universe is started at a sufficiently large value of
the scalar field $\phi$.

       One should  point out that the model discussed here gives a
potential with two physically relevant parameters
$\frac{f_1^{2}}{4f_{2}}$ , which represents
the value of $V_{eff}$ as $\phi \rightarrow \infty$ ,  i.e. the strength
of the false vacuum at
the
flat region and  $\frac{\alpha^{2}f_	{1}^{2}}{2f_2}$ , representing the
mass of the excitations around the
true vacuum with zero cosmological constant (achieved here without fine
tuning).

	When a realistic model of reheating is considered, one has to give
the strength of the coupling of the $\phi$ field to other fields. It
remains to
be seen what region of parameter space provides us with a realistic
cosmological model.

            Furthermore, one can consider this model as suitable for the
very late universe rather than for the early universe, after we suitably
reinterpret the meaning of the scalar field  $\phi$. 

	This can provide a long lived almost constant vacuum energy for a
long period of time, which can be small if $f_{1}^{2}/4f_{2}$ is
small. Such small energy
density will eventually disappear when the universe achieves its true
vacuum state. For a more detailed scenario which includes the effect of 
matter other than the dilaton  $\phi$ see next section.

	Notice that a small value of $\frac{f_{1}^{2}}{f_{2}}$   can be
achieved if we let $f_{2} >> f_{1}$. In this case
$\frac{f_{1}^{2}}{f_{2}} << f_{1}$, i.e. a very small scale for the
energy
density of the universe is obtained by the existence of a very high scale
(that of $f_{2}$) the same way as a small fermion mass is obtained in the
see-saw mechanism$^{7}$ from the existence also of a large mass scale. 

\section{Introducing Scale Invariant Mass Terms for Additional Matter
Fields and Cosmological Implications}

So far we have studied a theory which contains the metric tensor
$g_{\mu\nu}$, the measure fields $\varphi_{a}$ (a=1,2,3,4) and the
"dilaton" $\phi$, which makes global scale invariance possible in a
non-trivial way. All of the above fields have some kind of geometrical
significance, but if we are to describe the real world, the list of fields
and/or particles to be introduced has to be enlarged.

To see how scale invariant mass terms are possible, let us start with the
simplest possible example, i.e. the case of a point particle.

A point particle can be discussed by a contribution to $L_1$ and $L_2$ in
(9) of the form
\begin{equation}
L_{1p} = m_1 \int e^{-\alpha\phi/2} \sqrt{g_{\mu\nu}
\frac{dx^\mu}{d\lambda}
\frac{dx^\nu}{d\lambda}}
\frac{\delta^{(4)}(x-x(\lambda))}{\sqrt{-g}}d\lambda
\end{equation}
and
\begin{equation}
L_{2p} = m_2 \int e^{\alpha\phi/2} \sqrt{g_{\mu\nu}
\frac{dx^{\mu}}{d\lambda} \frac{dx^{\nu}}{d\lambda}}
\frac{\delta^{(4)}(x-x(\lambda))}{\sqrt{-g}} d\lambda.
\end{equation}

In this case, the contribution of $L_{1p}$ to the first term of (9) and of
$L_{2p}$ to the second term of (9) give rise to scale invariant
contributions under the transformations (14), (16), (17), (18) and (19).

Now, going through the same steps that lead us to the constraint (28), we
get now instead
\begin{equation}
M + V(\phi) - \frac{2U(\phi)}{\chi} + \frac{1}{2} (L_{1p} - \frac{1}{\chi}
L_{2p}) = 0
\end{equation}

If we are not located exactly on the particle, $\chi$ is given by the old
answer, i.e. $\chi$ = $2U(\phi)/(M + V(\phi))$. If, however, we are
located exactly at the point particle, the first three terms in (43) can
be ignored, since they are non-singular and we must then have
$L_{1p}$ - $\frac{1}{\chi} L_{2p}$ = 0, which means
\begin{equation}
\chi = \frac{m_2}{m_1} e^{\alpha\phi}.
\end{equation}

If an extended particle description of matter is taken, then (44) is
obtained in the region of high density of matter while $\chi$ =
$2U(\phi)/(M+V(\phi))$ is obtained for the low density of matter.

If (44) is inserted into (41), (42) and then both contributions of
$L_{1p}$ and $L_{2p}$ are inserted in (9), we obtain for the particle
contribution to the action
\begin{equation}
S_p = 2 \sqrt{m_1 m_2} \int \sqrt{\overline{g}_{\mu\nu} \frac{dx^\mu}{d\lambda}
\frac{dx^\nu}{d\lambda}} d\lambda
\end{equation}
where a transformation to the Einstein Frame $\overline{g}_{\mu\nu}$ =
$\chi g_{\mu\nu}$ has been made. We see than that no time dependent masses
are obtained, since in the Einstein Frame the $\phi$ dependence of the
particle action totally disappears.

Similar results are obtained if instead of a point particle, we use an
extended distribution of matter or a field, provided we use a high density
approximation.

Taking, for example, the case of a fermion $\psi$, where the kinetic term
of the fermion is chosen to be part of $L_1$
\begin{equation}
S_{fk} = \int L_{fk} \Phi d^4 x
\end{equation}
\begin{equation}
L_{fk} = \frac{i}{2} \overline{\psi} [\gamma^a V_a^\mu
(\overrightarrow{\partial}_\mu + \frac{1}{2} \omega_\mu^{cd} \sigma_{cd})
- (\overleftarrow{\partial}_\mu + \frac{1}{2} \omega_\mu^{cd} \sigma_{cd})
\gamma^a V^\mu_a] \psi
\end{equation}
there $V^\mu_a$ is the vierbein, $\sigma_{cd}$ =
$\frac{1}{2}[\gamma_c,\gamma_d]$, the spin connection $\omega^{cd}_\mu$ is
determined by variation with respect to $\omega^{cd}_\mu$ and, for
self-consistency, the curvature scalar is taken to be (if we want to deal
with $\omega_\mu^{ab}$ instead of $\Gamma^\lambda_{\mu\nu}$ everywhere)
\begin{equation}
R = V^{a\mu}V^{b\nu}R_{\mu\nu ab}(\omega),
R_{\mu\nu ab}(\omega)=\partial_{\mu}\omega_{\nu ab}
-\partial_{\nu}\omega_{\mu ab}+(\omega_{\mu a}^{c}\omega_{\nu cb}
-\omega_{\nu a}^{c}\omega_{\mu cb}).
\end{equation}

Global scale invariance (14), (16), (17), (18) and (19) is obtained
provided $\psi$ is also allowed to transform, as in
\begin{equation}
\psi \rightarrow \lambda ^{-\frac{1}{4}} \psi
\end{equation}

In this scale invariant case mass terms are of the form
\begin{equation}
S_{fm} = m_1 \int \overline{\psi} \psi e^{\alpha\phi/2} \Phi d^4x + m_2
\int \overline{\psi} \psi e^{3\alpha\phi/2} \sqrt{-g} d^4 x. 
\end{equation}

If we once again consider the situation where 
$m_1 e^{\alpha\phi/2} \overline{\psi}\psi$  
or $m_2 e^{3\alpha\phi/2} \overline{\psi}\psi$ are
much bigger than $V(\phi)$ + M, i.e. a high density approximation, we
obtain that instead of the constraint (28), the following holds,
\begin{equation}
(3m_2 e^{3\alpha\phi/2} + m_1 e ^{\alpha\phi/2} \chi) \overline{\psi}
\psi = 0
\end{equation}
which means
\begin{equation}
\chi = -\frac{3m_2}{m_1} e^{\alpha\phi}.
\end{equation}

Inserting (52) into (50), we obtain the $\phi$ independent mass term
after going to the conformal Einstein frame, which involves, when
fermions are present, also a transformation of the fermion fields,
necessary so as to achieve simultaneously the standard 
Einstein-Cartan form for both the gravitational and fermion 
equations. These transformations are, 
$\overline{g}_{\mu\nu}$ = $\chi g_{\mu\nu}$ (or
$\overline V_\mu^a$ = $\chi^\frac{1}{2} V_\mu^a$) and $\psi ^\prime$ =
$\chi ^{-\frac{1}{4}} \psi$ and they lead to a mass term,
\begin{equation}
S_{fm} = -2m_2 ( \frac {|m_1|}{3|m_2|})^{3/2}  \int\sqrt{-\overline {g}} 
\overline{\psi} ^{\prime} \psi ^{\prime} d^4x
\end{equation}

Once again, as in the case of the point particle, the $\phi$ dependence of
the mass term has disappeared (in this case, however, under the
approximation of high density of the fermion fields). The situation for low
density can be much more complicated and in fact could lead to a
non-conventional type of dynamics in this limit. Connections to proposals
for deviations from Newtonian dynamics appear possible, if one can
correlate low densities, where deviations are expected here, to low
accelerations, as is the case in Ref[8].

There is one situation where the low density of matter can also give results 
which are similar to those obtained in the high density approximation, in 
that the coupling of the $ \phi $ field disappears and that the mass term 
becomes of a conventional form in the Einstein conformal frame. For the point 
particle model this can never be the case, since the point particle always 
produces an infinite energy density at the point where it is located, but
such a discussion can be made in a meaningful way in the case of an extended
distribution, like that of a Dirac particle.

This is the case, when we study the theory for the limit 
$\phi \rightarrow \infty$ . Then $U(\phi) \rightarrow \infty$ and
$V(\phi) \rightarrow \infty$. In this case, taking
$m_1 e^{\alpha\phi/2} \overline{\psi}\psi$                                    
and $m_2 e^{3\alpha\phi/2} \overline{\psi}\psi$                                                                            
much smaller than $V(\phi)$ or $U(\phi)$ respectively and since 
also $M$ can be ignored in the constraint in this limit, we get then, 
 
\begin{equation}                                                              
\chi = \frac{2f_2}{f_1} e^{\alpha\phi}.                                       
\end{equation}                                                                
      
If (54) is inserted in (50), we get
\begin{equation}                                                              
S_{fm} = m \int\sqrt{-\overline {g}} 
\overline{\psi} ^{\prime} \psi ^{\prime} d^4x                                 
\end{equation}                                                                
 
where

\begin{equation}
 m = m_1(\frac {f_1}{2f_2})^{\frac{1}{2}} +  m_2(\frac {f_1}{2f_2})^{\frac{3}{2}}
\end{equation}

Comparing (55)-(56) and (53) and taking for example $m_1$ and $m_2$ of the same 
order of magnitude, we see that the mass of the Dirac particle is much 
smaller in the region $\phi \rightarrow \infty$, for which (55), (56) are 
valid, than it is in the region of high density of the Dirac particle 
relative to $V(\phi)+M$, as displayed in eq. (53), if the assumption
$\frac{f_1}{f_2} < < 1$, which was motivated in section 8, is made.

Therefore if space is populated by these diluted Dirac particles of this
type, the mass 
of these particles will grow substantially if we go to the true vacuum state 
valid in the absence of matter, i.e. $V+M=0$, as dictated by $V_{eff}$ given 
by eq. (36).

The presence of matter pushes therefore the minimum of energy to a state 
where $ V+M > 0$. The real vacuum in the presence of matter should not 
be located in the region $\phi \rightarrow \infty$, which minimizes the 
matter energy, but maximizes the potential energy $V_{eff}$ and not at
$V+M=0$, which minimizes $V_{eff}$, and where particle masses are big, but
somewhere in a balanced intermediate stage. Clearly how much above $V+M=0$
such true vacuum is located must be correlated to how much particle density 
is there in the Universe.     

The situation described by eq. (55)-(56) represents the situation of low
energy density of particles as compared to for example the false vacuum 
energy density, having there a very small mass. This mass can then grow
 a lot when the dilaton field approaches the minimum of its potential. 
This situation resembles
very much the scenario developed by Felder, Kofman and Linde$^{9}$, where, 
in the context of an inflationary scenario, particles, initially created 
with a very low mass, increase their mass considerably through the evolution 
of the inflaton field. This "fattening" of the particle masses can play a
role in making the transfer of energy from the inflaton field (in our case 
the dilaton field $\phi$ ) to matter very efficient.

Finally, let us mention that vector particles, even if massive, can be 
incorporated in a very simple way into the dilatation invariant theory 
described above. A scale invariant action, including mass is given by

\begin{equation}
S_{vector} = -\frac{1}{4} \int F_{\mu \nu} F^{\mu \nu}\sqrt{-g} d^4x +        
 \frac{m^2}{2} \int A_{\mu} A^{\mu} \Phi d^4 x 
\end{equation}

where
\begin{equation}
 F_{\mu \nu} =  \partial_{\mu}A_{\nu} -  \partial_{\nu}A_{\mu}
\end{equation}

Notice that if done in this way, mass  for the vector field  
$A_{\mu}$  is consistent with dilatation invariance, without need of a
coupling of such a field to the dilaton $\phi$.

\section{Discussion and Conclusions} 

In this paper we have seen that realistic realizations of scale invariance 
can be obtained when in addition to the standard measure of integration
$ \sqrt{-g}$, we also consider a measure of integration which is given
in terms of degrees of freedom which are independent of the metric. A 
dilaton field $\phi$ has to be introduced in order to make global scale 
invariance possible. Masses and potentials are possible in a way consistent
with scale invariance. This is achieved generically by allowing couplings
of fields to both possible measures. Then a non trivial dilaton potential 
appears which has attractive features from the point of view of cosmology, 
like an infinite region of flat potential which is desirable in new 
inflation.

Masses for other fields different from the dilaton can be obtained also
by coupling the mass terms to the two different measures. The coupling to
two different measures can be done even if we do not require scale invariance
and in this sense such idea can be exploited even outside the context of
scale invariant theories and this is of interest by itself. For the scale 
invariant case additional surprises appear. In the first place, for the high
density approximation, particles behave like regular particles and the 
coupling to the dilaton totally disappears  when we analyze the theory in 
the CEF. In some cases the low density
approximation of the fields can give also a normal propagation, i.e. standard
equations for the particles. This, as  we have seen, happens in the infinite 
flat region of the dilaton potential. 
In this region the mass is different to that obtained 
in the high density approximation, more reasonable for the region near the
true vacuum. Particles whose mass can naturally change and the cosmological
application of this in Felder, Kofman and Linde type scenarios in connection 
with the effective transfer of energy from the dilaton field to matter have 
been discussed. 

Also the fact that particle masses grow as we approach the state with zero 
cosmological constant can lead to an effect where the true vacuum in the
presence of matter is not the zero cosmological constant state (found to be 
the true vacuum in the absence of matter), but a state where $V_{eff} > 0$. 
How much above zero is $V_{eff}$, depends on the amount of particles present.

From the point of view of particle physics, this theory could give a new 
approach to the family problem, since here we have a situation where the same 
particle in different states can have a different, 
although well defined mass.

Finally the theory when applied to the study of  particles which are both 
low density and not in the infinite 
flat region of the potential could lead to a non conventional type of
dynamics, which could be related to that discussed by Bekenstein and Milgrom,
for example, if one can correlate low densities (needed here to get non 
conventional behavior) to low accelerations (as in the non conventional 
behavior of the models by Bekenstein and Milgrom).  

 \section{Acknowledgments}

I would like to thank J. Bekenstein, A. Davidson and A. Kaganovich for
conversations on
the subjects discussed here.

\break

\end{document}